\documentclass[journal]{IEEEtran}
\usepackage[nospace,noadjust]{cite}

\usepackage{amssymb,amsmath}
\usepackage{graphics}
\usepackage{float}
\usepackage{graphicx}
\usepackage{amsmath}
\usepackage{cite}
\usepackage{color}
\usepackage{dsfont}
\usepackage{bm}
\usepackage{upgreek}
\usepackage{arydshln}
\usepackage{enumerate}
\usepackage{amsthm}
\usepackage{graphicx}
\usepackage{threeparttable}
\usepackage{caption}
\usepackage{multirow}
\usepackage{overpic}
\usepackage{color}
\usepackage{xfrac}
\usepackage{array}
\usepackage{nomencl}
\usepackage{hyphenat}
\usepackage{scrextend}
\usepackage[monochrome]{xcolor}

\newcommand{\be}{\begin{equation}}
\newcommand{\ee}{\end{equation}}

\newcommand{\parder}[2]{ \frac{\partial #1}{\partial #2} }
\newcommand{\norm}[1]{ || #1 ||}
\newcommand{\mb}[1]{\mathbf{#1}}
\newcommand{\bs}[1]{\boldsymbol{#1}}
\newcommand{\virg}[1]{\textquotedblleft#1\textquotedblright}

\newcommand{\cvec}[1]{ \mathrm{vec}\left(  #1 \right) }
\newcommand{\tonde}[1]{\left( #1 \right)  }
\newcommand{\quadre}[1]{\left[  #1 \right]  }
\newcommand{\graffe}[1]{\left\lbrace   #1 \right\rbrace   }
\newcommand{\realpart}[1]{\mathrm{Re}\graffe{#1}}

\newtheorem{theorem}{Theorem}
\newtheorem{corollary}{Corollary}

\newtheorem{remark}{Remark}
\newtheorem{assumption}{Assumption}

\begin{document}
%
\title{Massive MIMO Radar for Target Detection}
%
%
%

\author{Stefano~Fortunati,~\IEEEmembership{Member,~IEEE,}
		Luca~Sanguinetti,~\IEEEmembership{Senior Member,~IEEE,}\\
        Fulvio~Gini,~\IEEEmembership{Fellow,~IEEE,}
                Maria~S.~Greco,~\IEEEmembership{Fellow,~IEEE,}
                Braham~Himed,~\IEEEmembership{Fellow,~IEEE}\vspace{-0.6cm}
\thanks{S.~Fortunati, L.~Sanguinetti, F.~Gini and M.~S.~Greco are with University of Pisa, Dipartimento di Ingegneria dell'Informazione, Via Caruso, 56122, Pisa, Italy. \newline\indent  Braham Himed is with  Air Force Research Laboratory, Dayton OH, USA. \newline\indent  This work has been partially supported by the Air Force Office of Scientific Research under award number FA9550-17-1-0344.} }

\maketitle


\begin{abstract}
	Since the seminal paper by Marzetta from 2010, the Massive MIMO paradigm in communication systems has changed from being a theoretical scaled-up version of MIMO, with an infinite number of antennas, to a practical technology. Its key concepts have been adopted in the 5G new radio standard and base stations, where $64$ fully-digital transceivers have been commercially deployed. Motivated by these recent developments, this paper considers a co-located MIMO radar with $M_T$ transmitting and $M_R$ receiving antennas and explores the potential benefits of having a large number of virtual spatial antenna channels $N=M_TM_R$. Particularly, we focus on the target detection problem and develop a \textit{robust} Wald-type test that guarantees certain detection performance, regardless of the unknown statistical characterization of the disturbance. Closed-form expressions for the probabilities of false alarm and detection are derived for the asymptotic regime $N\to \infty$. Numerical results are used to validate the asymptotic analysis in the finite system regime with different disturbance models. Our results imply that there always exists a sufficient number of antennas for which the performance requirements are satisfied, without any a-priori knowledge of the disturbance statistics. This is referred to as the Massive MIMO regime of the radar system.\end{abstract}


\begin{IEEEkeywords}
Large-scale MIMO radar, robust detection, Wald test, misspecification theory, unknown disturbance distribution, dependent observations.
\end{IEEEkeywords}


\vspace{-0.2cm}
\section{Introduction}


Consider a multiple antenna radar system characterized by $N$ spatial channels collecting $K$ temporal snapshots $\{\mb{x}_k\}_{k=1}^K \in\mathbb{C}^N$ from a specific resolution cell, defined in an absolute reference frame. The primary goal of any radar system is to discriminate between two alternative hypotheses: the presence ($H_1$) or absence ($H_0$) of the target, in the resolution cell under test. Among others, a common model for the signal of interest is $\bar{\alpha}_k \mb{v}_k$, where $\mb{v}_k\in\mathbb{C}^N$ is known at each time instant $k \in \{1,\ldots,K\}$ and $\bar{\alpha}_k\in\mathbb{C}$ is a deterministic, but \textit{unknown}, scalar that may vary over $k$. Any measurement process involves a certain amount of disturbance. In radar signal processing, the disturbance is produced by two components, the \emph{clutter} and white Gaussian measurement noise, and it is modelled as an additive random vector, say $\mb{c}_k$, whose statistics may vary over $k$. Formally, the detection problem can be recast as a \textit{composite} binary hypothesis test (HT) \cite{STAP}:
\be
\label{HT_base_1}
\begin{array}[l]{clc}
	H_0: & \mb{x}_k = \mb{c}_k &  k = 1,\ldots,K, \\
	H_1: & \mb{x}_k = \bar{\alpha}_k \mb{v}_k+ \mb{c}_k & k = 1,\ldots,K.
\end{array}
\ee
To solve \eqref{HT_base_1}, a decision statistic $\Lambda(\mb{X})$ of the dataset $\mb{X} \triangleq [\mb{x}_1,\ldots,\mb{x}_K]$ is needed and its value must be compared with a threshold: 
\begin{align}\label{eq:test}
\Lambda(\mb{X}) \overset{H_1}{\underset{H_0}{\gtrless}} \lambda
\end{align}
to discriminate between the null hypothesis $H_0$ and the alternative $H_1$. A common requirement in radar applications is that the probability of false alarm has to be maintained below a pre-assigned value, say $ \overline{P_{FA}}$. Consequently, the threshold $\lambda$ should be chosen to satisfy the following integral equation:
\be
\label{int_eq_1}
\Pr \graffe{\Lambda(\mb{X}) > \lambda | H_0}=\int_{\lambda}^{\infty} p_{\Lambda|H_0}(a|H_0)da =  \overline{P_{FA}},
\ee           
where $p_{\Lambda|H_0}$ is the probability density function (pdf) of $\Lambda(\mb{X})$ under the null hypothesis $H_0$.
\vspace{-0.2cm}
\subsection{Motivation}

Finding a solution to \eqref{int_eq_1} is in general a challenge. The common way out relies upon some \virg{ad-hoc} assumptions on the statistical model of the dataset $\mb{X}$. In order to clarify this point, let us have a closer look to the steps required to solve \eqref{int_eq_1}. Firstly, a closed-form expression for $p_{\Lambda|H_0}$ is needed. By definition, $p_{\Lambda|H_0}$ is a function of the chosen decision statistic $\Lambda(\mb{X})$ and of the joint pdf $p_{\bs{X}}(\mb{X})$ of $\mb{X}$. If all the $\bar{\alpha}_k, \forall k$ are modelled as deterministic unknown scalars, $p_{\bs{X}}(\mb{X})$ is fully determined by the joint pdf $p_{\bs{C}}(\mb{C})$ of the disturbance $\mb{C} = [\mb{c}_1,\ldots,\mb{c}_K]$. A first simplification comes from the assumption that the disturbance vectors $\{\mb{c}_k\}$ are independent and identically distributed (i.i.d.) random vectors such that $p_{\bs{C}}(\mb{C}) = \prod\nolimits_{k=1}^{K}p_{C}(\mb{c}_k)$ \cite{STAP_IID_ass_1,STAP_IID_ass_2}. This assumption is, however, not always valid in practice. A second simplification that is commonly adopted in the radar literature (see e.g. \cite{Himed_GLRT,STAP_IID_ass_1,STAP_IID_ass_2}) is to assume that the functional form of $p_{C}(\mb{c}_k) \equiv p_{C}(\mb{c}_k; \bs{\gamma})$ is perfectly known, up to a possible (finite-dimensional) deterministic nuisance vector parameter $\bs{\gamma}$; for example,  the (vectorized) covariance matrix. In order to obtain a consistent estimate $\hat{\bs{\gamma}}$ of $\bs{\gamma}$, a \textit{secondary dataset}\footnote{In radar terminology, a secondary dataset is a set of \virg{signal-free} snapshots collected form resolution cells adjacent to the one under test and sharing the same statistical characterization.} has to be exploited (see e.g. \cite{Ward}). Note that the required $p_{\Lambda|H_0}$ is a function of $\hat{\bs{\gamma}}$ as well. A third simplifying assumption is that the signal parameters $\bar{\alpha}_k$ remain constant over $k$, i.e. $\bar{\alpha}_k \equiv \bar{\alpha}, \forall k$ \cite{Himed_GLRT,STAP_IID_ass_1,STAP_IID_ass_2}. Under these three assumptions, a possible choice for the decision statistic is the generalized likelihood ratio (GLR) $\Lambda_{\mathsf{GLR}}(\mb{X})$ (see e.g. \cite[Ch. 11]{kay1993fundamentalsII} and references therein). However, a closed-form solution to \eqref{int_eq_1} can be found only for a very limited class of disturbance models for which the Gaussianity assumption needs also to be imposed. An asymptotic approximation for the solution to \eqref{int_eq_1} can be obtained by exploiting a well-known asymptotic property of the GLR. Under the hypothesis $H_0$ and for $K\to \infty$, the pdf of $\Lambda_{\mathsf{GLR}}(\mb{X})$ converges to the one of a central $\chi$-squared random variable with $2$ degrees of freedom, denoted as $\chi_2^2(0)$ \cite[Ch. 11]{kay1993fundamentalsII}. Hence, by using the properties of the $\chi$-squared distribution, it is immediate to verify that
\eqref{int_eq_1} is asymptotically satisfied by $\bar{\lambda} = -2\ln \overline{P_{FA}}$. This is a particularly simple result that has received a lot of attention in the literature. However, it relies on the four simplifying assumptions previously introduced and summarized as follows:
\begin{itemize}
	\item[A1] The disturbance vectors $\{\mb{c}_k\}_{k=1}^K$ are i.i.d. over the observation interval.
	\item[A2] The pdf $p_{C}(\mb{c}_k; \bs{\gamma})$ of the disturbance is perfectly known, up to a unknown nuisance parameter vector $\bs{\gamma}$.
	\item[A3] The target complex amplitude $\bar{\alpha}_k$ is maintained constant over the observation interval, i.e. $\bar{\alpha}_k \equiv \bar{\alpha}, \forall k$.
	\item[A4] \textcolor{blue}{The number of temporal snapshots $K$ is assumed to be much larger that the spatial channels $N$.}
\end{itemize}
Even if these assumptions make the (asymptotic) analysis of $\Lambda_{\mathsf{GLR}}(\mb{X})$ analytically tractable, they are seldom satisfied in practical applications.

\subsection{Contributions}

This paper considers a co-located MIMO radar with $M_T$ transmitting and $M_R$ receiving antennas and aims at deriving a detector that satisfies pre-assigned performance requirements without relying on the four assumptions above. Inspired by the recent developments in Massive MIMO communications \cite{marzetta2010noncooperative,BjornsonHS17,massivemimobook,Bjornson2019a}, we aim at exploring the potential benefits of having a very large number of antennas. Particularly, we assume that a single time snapshot, i.e. $K=1$, is collected, and operate in the asymptotic regime where the number of virtual spatial antenna channels $N = M_TM_R$ grows unboundedly, i.e., $N\to\infty$. {\color{blue}This makes the three assumptions A1, A3 and A4 no longer needed.} Advances in robust and misspecified statistics (\cite{Huber,white,SPM,TSP_MCRB,Con_MCRB,miss_sb} and \cite{white_nl_reg_dep, MMLE_dep}) are used to dispose of the cumbersome and unrealistic assumption A2.
By adopting a very general disturbance model taking into account the spatial correlation structure of the observed samples, we propose a robust Wald-type detector that is asymptotically distributed, when $N\to \infty$, as a $\chi$-squared random variable (under both $H_0$ and $H_1$) irrespective of the actual and  unknown disturbance pdf $p_{C}(\mb{c})$. This asymptotic result is achieved without the need of any secondary dataset. Although the theoretical findings of this paper are valid for a very general disturbance model, numerical results are provided for two non-Gaussian, stable auto-regressive disturbance models of order $p=3$ and $6$. It turns out that a pre-assigned value of $\overline{P_{FA}}=10^{-4}$ is achieved for $N= M_TM_R\ge 10^4$ with both models. This number of virtual spatial antenna channels defines what we call the \emph{Massive MIMO regime} of the radar system. 

\textcolor{blue}{Compared to our previous paper \cite{icassp}, the main difference lies in the absence of any a priori knowledge of the disturbance model. In fact, in \cite{icassp} the analysis was developed for an autoregressive model of order $1$, but with no a priori knowledge of its statistics.}
%


\subsection{Relevant literature}

The MIMO paradigm has been the subject of intensive research over the past 15 years in radar signal processing. Initially introduced in wireless communications as a new enabling technology, the MIMO framework has been recognized to have a great potential in boosting the capabilities of classical antenna array systems. Based on the array configurations used, MIMO radars can be classified into two main types. The first type uses widely separated antennas (so-called distributed MIMO) to capture the spatial diversity of the target's radar cross section (RCS) \cite{wide_MIMO}. The second type employs arrays of closely spaced antennas (so-called co-located MIMO) to coherently combine the probing signals in certain points of the search area \cite{Stoica_col}. Hybrid configurations are also possible.

While the advantages in terms of spatial resolution, parameter identifiability, direction-of-arrival estimation and interference mitigation have been largely investigated in the MIMO radar literature, the potential benefits that a large number of virtual spatial antenna channels can bring into the target detection problem in terms of robustness with respect to the generally unknown disturbance model have not been explored yet. Surprisingly, not only the highly desirable robustness property has been somehow disregarded but, as pointed out in \cite{Fried_7}, even the availability of reliable, non-trivial, disturbance models is scarce. Remarkable exceptions to the mainstream Gaussianity assumption have been recently discussed in \cite{Abla} and in \cite{Large_scale}. Particularly, in \cite{Abla} the performance of the Adaptive Normalized Matched Filter (ANMF), exploiting robust estimators for the disturbance covariance matrix, has been investigated with non-Gaussian disturbance. Specifically, random matrix tools have been used to obtain asymptotic approximations of the probabilities of false alarm and misdetection of the ANMF for the regime in which both $N$ and $K$ go to infinity with a non-trivial ratio $N/K$. Similar random matrix tools have been adopted in \cite{Large_scale} to derive some asymptotic (in random matrix regime) results about the direction-of-departure and direction-of-arrival estimation in a non-Gaussian disturbance setting. Again, the random matrix machinery has been exploited in \cite{Debbah} to investigate the asymptotic performance of a GLRT detector in cognitive radio applications. Specifically, the HT problem tackled in \cite{Debbah} is similar to the one in \eqref{HT_base_1}, but the steering vector $\mb{v}_k$ and $\bar{\alpha}_k$ are assumed unknown. \textcolor{blue}{In the same spirit of \cite{Debbah}, \cite{kob} has recently investigated the possibility to derive eigenvalues-based detectors for HT problem of the form in \eqref{HT_base_1} for spectrum sensing and sharing in cognitive radio.} However, in both \cite{Debbah} and \cite{kob}, the disturbance was assumed to be a simple white Gaussian process with distribution a priori known, up to its statistical power. The asymptotic analysis in \cite{Large_scale, Debbah,kob} requires that both $N$ and $K$ grow unboundedly. This is different from this paper where the temporal dimension $K$ is kept fixed; specifically, we assume to collect a single snapshot vector. 

\subsection{Outline and Notation} The reminder of the paper is organized as follows. \textcolor{blue}{In Section II, the system and signal models as well as the resulting HT problem are introduced, by focusing our attention on the disturbance model for co-located MIMO radars. Section III describes our main results. Specifically, the explicit form of the proposed robust Wald-type test is provided along with the theoretical derivation of its asymptotic distribution under both $H_0$ and $H_1$. Numerical validations and simulation results are provided in Section IV. Some concluding remarks and discussions are drawn in Section V}.

Throughout this paper, italics indicates scalar quantities ($a$), lower case and upper case boldfaces indicate column vectors ($\mathbf{a}$) and matrices ($\mathbf{A}$), respectively. Each entry of a $N \times N$ matrix $\mb{A}$ is indicated as $a_{i,j}\triangleq [\mb{A}]_{i,j}$, while the $i$-th column vector of $\mb{A}$ is indicated as $\mb{a}_i$ such that $\mb{A} = [\mb{a}_1,\ldots,\mb{a}_N]$. We use $^*$, $^T$, and $^H$ to indicate complex conjugation, transpose, and the Hermitian operators, respectively. For random  variables or vectors, $=_d$ stands for \virg{has the  same  distribution  as}. Also, $\overset{a.s.}{\underset{N\rightarrow \infty}{\rightarrow}}$ indicates the almost sure (a.s.) convergence and $\overset{p}{\underset{N\rightarrow \infty}{\rightarrow}}$ indicates the convergence in $p$-probability. We call $Q_1(\cdot,\cdot)$ the Marcum $Q$ function of order 1. The symbol $\lfloor a \rfloor$ defines the nearest integer less than or equal to $a \in \mathbb{R}$.


\section{System Model and Problem Formulation}
\label{sect_II}
\begin{figure}[t!]
	\centering
	\begin{overpic}[width=0.8\columnwidth]{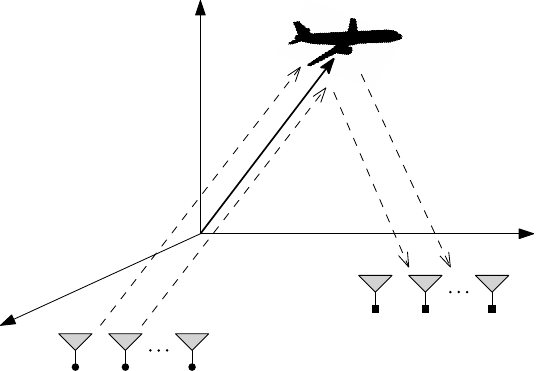}
		\put(58,68){Target, $\bs{\phi}$}
		\put(30,8){Transmitter}
		\put(55,20){Receiver}
	\end{overpic}
	\caption{Co-located MIMO radar.\vspace{-0.5cm}}
	\label{fig:Fig1}
\end{figure}

Consider a co-located MIMO radar system equipped with $M_T$ transmitting antennas and $M_R$ receiving antennas \cite{Stoica_col}. The transmitting array is characterized by the \textit{array manifold}, also called steering vector, $\mb{a}_T(\bs{\phi})$, where $\bs{\phi}$ is the position vector defined in an absolute reference frame \cite{Stoica_MIMO_1}. Similarly, the receiving array can be characterized by the steering vector $\mb{a}_R(\bs{\phi})$ since the positions of the antennas in the absolute reference frame are known; see Fig. \ref{fig:Fig1}.

\subsection{Signal model}
Given a target with position vector $\bar{\bs{\phi}}$, the signal collected at the receiving array can be modelled as \cite{Fried,Fried_7}:
\be
\label{cont_time_model}
\mb{x}(t) = \bar{\alpha}\mb{a}_R(\bar{\bs{\phi}})\mb{a}_T^T(\bar{\bs{\phi}}) \mb{s}(t-\bar{\tau})e^{\mathsf{j}\bar{\omega}t} + \mb{n}(t), \; t \in [0,T],
\ee  
where $\mb{x}(t) \in \mathbb{C}^{M_R}$ is the array output vector at time $t$, $\mb{s}(t) \in \mathbb{C}^{M_T}$ is the vector of transmitted signals, $\bar{\alpha} \in \mathbb{C}$ accounts for the radar cross section of the target and the two-way path loss, which is the same for each transmitter and receiver pair. This is generally verified in co-located MIMO radars \cite{Stoica_col}. The parameters $\bar{\tau}$ and $\bar{\omega}$ represent the actual time delay and Doppler shift, due to the target position and velocity. The complex, vector-valued, random process $\mb{n}(t) \in \mathbb{C}^{M_R}$ accounts for the disturbance. We assume that $\mb{s}(t)$ is obtained as a linear transformation of a set of \textit{nearly} orthonormal
signals $\mb{s}_o(t)\in \mathbb{C}^{M_T}$, i.e. $\mb{s}(t) = \mb{W}\mb{s}_o(t)$, where $\mb{W} = [\mb{w}_1,\ldots,\mb{w}_{M_T}]^T \in \mathbb{C}^{M_T \times M_T}$ and $\mb{w}_m \in \mathbb{C}^{M}$ is the weighting vector of the transmit antenna element $m$ with power $\norm{\mb{w}_m}^2$.

\textcolor{blue}{Let $l=1,\ldots,L$ and $k=1,\ldots,K$ be the indices characterizing each time $l\Delta t$ and frequency $k\Delta \omega$ samples, respectively}. The output $\mb{X}(l,k)\in \mathbb{C}^{M_R \times M_T}$ of the linear filter matched to $\mb{s}_o(t)$ can be expressed as \cite{Fried,Fried_7}:
\be
\label{sig_mod}
\begin{split}
	\mb{X}&(l,k) = \bar{\alpha} \mb{a}_R(\bar{\bs{\phi}})\mb{a}_T(\bar{\bs{\phi}})^T \mb{W} \mb{S}(l,k) + \mb{C}(l,k),
\end{split}
\ee
where
\be\label{S_mat}
\mb{S}(l,k) \triangleq \int_{0}^{T} \mb{s}_o(t-\bar{\tau})\mb{s}_o^H(t-l\Delta t)e^{-j(k\Delta\omega-\bar{\omega})t}dt
\ee
takes into account potential \virg{straddling losses}, that are losses due to a not precise centering of the target in a range-Doppler gate	
or to a not exact orthogonality between waveforms, and
\be\label{n_MF}
\mb{C}(l,k)= \int_{0}^{T} \mb{n}(t)\mb{s}_o^H(t-l\Delta t)e^{-jk\Delta \omega t}dt.
\ee
After omitting the indexes $(l,k)$ for ease of notation, we may rewrite \eqref{sig_mod} in vectorial form as:
\be
\label{vec_data_model}
\mathbb{C}^{N} \ni \mb{x} = \cvec{\mb{X}} = \bar{\alpha} \mb{v}(\bar{\bs{\phi}}) + \mb{c},
\ee 
where $N \triangleq M_R M_T$ and 
\be\label{v_vect}
\mb{v}(\bar{\bs{\phi}}) = \left(\mb{S}^T \otimes \mb{I}_{M_R}\right)\left[{\mb{W}^T\mb{a}_T(\bar{\bs{\phi}}) \otimes \mb{a}_R(\bar{\bs{\phi}})}\right]
\ee
and $\mb{c} \triangleq \cvec{\mb{C}}$. {\color{blue}While $\mb{a}_T(\bs{\phi})$ and $\mb{a}_R(\bs{\phi})$ depend on the geometry of the transmitting and receiving arrays, respectively, the matrix $\mb{W}$ can be designed to shape arbitrarily the transmitting beam; see \cite[Ch. 4]{libro_W} and references therein. For example, with $\mb{W}=\mb{I}_{M_T}$, the transmitted power is uniformly distributed over all possible directions. On the other hand, with $\mb{W}=\mb{a}_T(\bs{\phi})^*\mb{a}_T(\bs{\phi})^T$, it is fully directed towards the direction $\bs{\phi}$. Intermediate cases can be obtained \cite{Fried}.} 

We assume that $\mb{n}(t)$ is zero-mean and wide-sense stationary; that is, $E\{\mb{n}(t)\} = \mb{0}$, $\forall t$ and $E\{\mb{n}(t)\mb{n}(\tau)^H\} = \bs{\Sigma}(t-\tau)$. Hence, from \eqref{n_MF} it easily follows that $E\{\mb{c}\} = \mb{0}$ and
\begin{align}\notag \label{C_int}
\bs{\Gamma}\triangleq E\{\mb{c}\mb{c}^H\} &= \int_{0}^{T} \int_{0}^{T}  \quadre{\mb{s}_o^*(t-l\Delta t) \otimes \mb{I}_{M_R}}  \mb{\Sigma}(t-\tau) \times \\ \notag
& \times \quadre{\mb{s}_o^*(t-l\Delta t) \otimes \mb{I}_{M_R}}^He^{-jk\Delta \omega (t-\tau)}dtd\tau \\\notag
&=\int_{0}^{T} \int_{0}^{T}  \quadre{\mb{s}_o^*(t-l\Delta t)\mb{s}_o^T(t-l\Delta t) \otimes \bs{\Sigma}(t-\tau)} \times\\
& \times e^{-jk\Delta \omega (t-\tau)}dtd\tau.
\end{align}
As seen, $\bs{\Gamma}$ is a function of $\bs{\Sigma}(t-\tau)$ (i.e., the covariance matrix of $\mb{n}(t)$) and $\mb{s}_o(t)$. In the literature (e.g., \cite{MIMO_TAB},\cite[Ch. 4]{Stoica_book_MIMO}), a simple model for $\mb{n}(t)$ is to assume that its samples are uncorrelated in both spatial (along the receiving array) and temporal (along $T$) domains. This implies that $\bs{\Sigma}(t-\tau) = \sigma^2\mb{I}_{M_R}\delta(t-\tau)$. If $\mb{s}_o(t)$ is a vector of orthonormal waveforms, it is thus immediate to verify that $\bs{\Gamma}$ in \eqref{C_int} reduces to $\bs{\Gamma} = \sigma^2\mb{I}_{N}$. Under the assumption of uncorrelated samples in the time domain only and perfect orthonormality of the transmitted waveforms, we have that $\bs{\Gamma} = \mb{I}_{M_T} \otimes \bs{\Sigma}_R$ where $\bs{\Sigma}_R$ denotes the receive spatial covariance matrix \cite{Stoica_Q_2}. However, the above two conditions may not be satisfied in practice \cite{Fried_non_ort, Fried_7}. This is why in this paper we do not make any a priori assumption on the structure of $\bs{\Gamma}$. We only assume that its $(i,j)$-th entry goes to zero at least polynomially fast as $|i-j|$ increases. 
This assumption will be discussed in the next section and formally introduced in Assumption \ref{assumption_mix}. 


\subsection{Disturbance model}\label{c_model}
As previously discussed, many simplified models have been proposed in the literature to statistically characterize the disturbance vector $\mb{c}$ at the output of the matched filter bank of a (co-located) MIMO radar system. We refer to \cite{Fried_non_ort,Fried_7} for a comparison among various statistical models and for a discussion about the physical simplifying assumptions underlying them. Here, we simply note here that the two main hypotheses usually made about the statistical characterization of the disturbance vector are: \textit{i}) $\mb{c}$ is temporally and spatially white, and \textit{ii}) $\mb{c}$ is Gaussian-distributed. \textcolor{blue}{ As we will show, advances in robust and mis-specified statistics allow us to drop these two strong assumptions in favor of much weaker conditions.}

To formally characterize the class of random processes to which the results of this paper apply, we need to introduce the concepts of \textit{uniform} and \textit{strong mixing} random sequences \cite{bradley2005,book_CLT_dep_2,book_CLT_dep_1,CLT_dep}. Roughly speaking, the uniform and strong mixing properties characterize the dependence between two random variables extracted from a discrete process separated by $m$ lags. Without any claim of completeness or measure-theoretic rigor, the results of this paper apply to any random process that satisfies a restriction on the speed of decay of its auto-correlation function. Specifically, we limit ourselves to the following class of random processes:
\begin{assumption}\label{assumption_mix}
	Let $\{c_n: \forall n\}$ be a stationary discrete and circular complex-valued process \cite{Pici} representing the true, and generally unknown, disturbance. Then, we assume that its autocorrelation function satisfies $r_C[m] \triangleq E\{c_nc^*_{n-m}\}=O(|m|^{-\gamma})$, $m \in \mathbb{Z}$, $\gamma > \varrho/(\varrho-1)$, $\varrho>1$.\footnote{Given a real-valued function $f(x)$ and a positive real-valued function $g(x)$, $f(x)=O(g(x))$ if and only if there exists a positive real number $a$ and a real number $x_0$ such that $|f(x)|\leq a g(x),\; \forall x \geq x_0$.}
\end{assumption} 
Assumption \ref{assumption_mix} implies that the volume of the MIMO radar must increase with the number of transmitting ($M_T$) and receiving ($M_R$) antennas. This means that the results of this paper do not apply to \virg{space-constrained} array topologies. That said, Assumption \ref{assumption_mix} is very general and allows to account for most practical disturbance models. Indeed, any (not necessary Gaussian) stable \textit{second-order stationary} (SOS) $\mathsf{ARMA}(p,q)$, and consequently any stable SOS $\mathsf{AR}(p)$ \cite{Pici_AR}, satisfies Assumption \ref{assumption_mix}, since the auto-correlation function  of any stable SOS ARMA decays exponentially. The generality of the ARMA model is because it can approximate, for $p$ and $q$ sufficiently large, the second-order statistics of any complex discrete random processes having a \textit{continuous} Power Spectral Density (PSD) \cite[Ch. 3]{Stoica_book}. Moreover, a non necessarily Gaussian $\mathsf{ARMA}(p,q)$ is able to model the \virg{spikiness} of heavy-tailed data as well. Another disturbance model of practical interest satisfying Assumption \ref{assumption_mix} is the Compound-Gaussian (CG) model \cite{CG_ran,Sang}. Indeed, recall that any CG-distributed random vector 
$\mb{c}$ admits a representation:
\be\label{CG_c}
\mb{c}=_d \sqrt{\tau}\mb{m}
\ee    
for some real-valued positive random variable $\tau$, called \textit{texture}, independent of the zero-mean, $N$-dimensional, circular, complex Gaussian random vector, called \textit{speckle}, $\mb{m} \sim \mathcal{CN}(\mb{0},\bs{\Gamma})$, where $\bs{\Gamma}$ is its scatter matrix. Under a condition on the structure of $\bs{\Gamma}$, it easily follows that the $N$ entries of $\mb{m}$ can be interpreted as $N$ random variables extracted from a circular, Gaussian, SOS $\mathsf{ARMA}(p,q)$ process $\{m_n: \forall n \}$, with $p,q<N$.

\textcolor{blue}{We conclude by noticing that Assumption \ref{assumption_mix} is more general than the one adopted in our previous work \cite{icassp}. In fact, the asymptotic results derived in \cite{icassp} are obtained by assuming an AR disturbance model,\footnote{We considered only the $\mathsf{AR}(1)$ case, but the theory could be readily extended to a general $\mathsf{AR}(p)$ as discussed in \cite{MMLE_time_series}.} driven by innovations with possibly unknown pdf. Unlike \cite{icassp}, this paper does not require any a priori information on the specific disturbance model; as stated in Assumption \ref{assumption_mix}, only the polynomial decay of its auto-correlation function is needed. 
} 

\subsection{The hypothesis test problem}\label{HT_model}
Based on the assumptions discussed above, the HT problem for target detection in \eqref{HT_base_1} can be expressed as:
\be
\label{HT_base_x}
\begin{array}[l]{clc}
	H_0: & \mb{x} = \mb{c},  \\
	H_1: & \mb{x} = \bar{\alpha} \mb{v}+ \mb{c},
\end{array}
\ee
where $\mb{c}\in\mathbb{C}^N$ is the disturbance vector whose entries are sampled from a complex random process $\{c_n: \forall n\}$ satisfying Assumption \ref{assumption_mix}. Note that, in practical radar scenarios, \eqref{HT_base_x} needs to be solved for any radar resolution cell of interest. Specifically, let $\{\bs{\phi}_i;i=1,\ldots,Q\}$ be the set of position vectors pointing at $Q$ radar resolution cells, defined in an absolute reference frame. Then, the presence (or absence) of a target has to be tested for all the $Q$ resolution cells. Consequently, \eqref{HT_base_x} has to be solved for any different vector $\mb{v}(\bs{\phi}_i)$, whose explicit form is defined in \eqref{v_vect}. Notice also that a single-snapshot, i.e., $K=1$, is used in \eqref{HT_base_x}. 

\textcolor{blue}{To discriminate between $H_0$ and $H_1$ in the composite HT problem \eqref{HT_base_x}, a test statistic is needed. Classical \textit{model-based} test statistics such as the Generalized Likelihood Ratio (GLR) test and the Wald test (see e.g. \cite{Wald_de_maio_2}) cannot be used since the functional form of the disturbance pdf $p_C$ and, consequently, of the data pdf $p_X$ are unknown. We propose the following approach. \textit{Since no a priori information on the functional form of $p_X$ is available, let us choose the simplest estimator, say $\hat{\alpha}$, for the signal parameter $\bar{\alpha}$. Then, building upon the asymptotic statistics of $\hat{\alpha}$ as $N \rightarrow \infty$, we define a Wald-type test to solve the HT problem in \eqref{HT_base_x}}. In fact, unlike the GLR statistic that requires the explicit functional form of the data pdf, a Wald-type statistic only needs an asymptotically normal, $\sqrt{N}$-consistent estimator of $\bar{\alpha}$ and a consistent estimate of its error covariance. As shown next, this key fact allows us to derive a \textit{robust} test statistic for Massive MIMO radar configurations.}



\section{Main Results}
\label{main}
The main result of this section is the derivation of a robust Wald-type test for \eqref{HT_base_x} with the valuable property to have, under $H_0$, an asymptotic distribution invariant w.r.t. $p_X$. The closed form expressions of its asymptotic distribution under both $H_0$ and $H_1$ will be provided. Since \eqref{HT_base_x} is a composite HT problem (i.e., it depends on the unknown deterministic signal parameter $\bar{\alpha}$) a prerequisite for the implementation of a decision statistic is the derivation of an estimator $\hat{\alpha}$ of  $\bar{\alpha}$.   

\subsection{Estimation of $\bar{\alpha}$ under dependent data}



By relying on the outcomes of \cite{MMLE_dep,white_nl_reg_dep}, in the following we show that an asymptotically normal, $\sqrt{N}$-consistent estimator of $\bar{\alpha}$ can be easily implemented under any general disturbance models satisfying Assumption \ref{assumption_mix}.  

A standard procedure is to recast an estimation problem into a relevant (possibly constrained) optimization problem. In the application at hand, an estimate of ${\alpha}$ can be obtained as:
\be
\label{MMLE}
\hat{\alpha} =  \underset{\alpha \in \mathbb{C}}{\mathrm{argmin}} \; G_N(\mb{x},\alpha),\quad \mb{x} \sim p_{X},
\ee
where $G_N(\cdot,\cdot)$ is a suitable objective function and $\mb{x}=[x_1,\ldots,x_N]^T \in \mathbb{C}^N$ is the available dataset characterized by an unknown joint pdf $p_{X}$.


Given the measurement model (under $H_1$) in \eqref{HT_base_x}, the most natural choice for $G_N(\cdot,\cdot)$ is a Least-Square (LS) based objective function:
\be
\label{obj_fun}
\begin{split}
	G_N&(\mb{x},\alpha) \triangleq \sum_{n=1}^{N} |x_n - \alpha v_n|^2 = (\mb{x}-\alpha\mb{v})^H(\mb{x}-\alpha \mb{v})\\
	& = \norm{\mb{x}}^2 + \norm{\mb{v}}^2\left|\alpha - \frac{\mb{v}^H\mb{x}}{\norm{\mb{v}}^2} \right|^2 - \frac{|\mb{v}^H\mb{x}|^2}{\norm{\mb{v}}^2} 
\end{split}
\ee
whose minimum is reached when the second addend vanishes. This yields  
\be
\label{alpha_est}
\hat{\alpha} = \frac{\mb{v}^H\mb{x}}{\norm{\mb{v}}^2},
\ee
which is a well-known result in the radar signal processing literature addressing decision rules in Gaussian environment. It can also be noted that, under a misspecified (white) Gaussian assumption on the disturbance vector $\mb{c}$, the LS estimator in \eqref{alpha_est} coincides with the Mismatched Maximum Likelihood estimator \cite{SPM},\cite{MMLE_dep}. 

By specializing the general findings about misspecified estimation under dependent data proposed in \cite{white_nl_reg_dep} and \cite{MMLE_dep}, the asymptotic properties of the LS estimator in \eqref{alpha_est} can be obtained as shown in the following Theorem \ref{MMLE_asyn}.   
%
%

\begin{theorem}
	\label{MMLE_asyn}
	Under Assumption \ref{assumption_mix}, the LS estimator $\hat{\alpha}$ in \eqref{alpha_est} is $\sqrt{N}$-consistent
			\be
		\hat{\alpha}\overset{a.s.}{\underset{N\rightarrow \infty}{\rightarrow}} \bar{\alpha}
		\ee
		and asymptotically normal
				\be
		\label{asym_norm}
		\sqrt{N}\bar{B}_N^{-1/2}A_N\times (\hat{\alpha} - \bar{\alpha}) \underset{N\rightarrow \infty}{\sim} \mathcal{CN}(0,1),
		\ee
		where 
		\begin{align}
		\label{A_def}
		A_N &\triangleq N^{-1}\norm{\mb{v}}^2,\\
		\label{B_def}
		\bar{B}_N &\triangleq N^{-1}{\mb{v}^H\bs{\Gamma}\mb{v}},
		\end{align}
		and $\bs{\Gamma} \triangleq E_{p_C}\{\mb{c}\mb{c}^H\}$, with $p_C$ being the true (but generally unknown) disturbance pdf. 
\end{theorem}

\begin{IEEEproof}
All the required regularity conditions and a measure-theoretic rigorous proof (for the real-valued case) can be found in \cite{white_nl_reg_dep} and \cite{MMLE_dep}, while in Appendix A of this paper we provide the reader with an \virg{easy-to-follow} but still insightful sketch of the proof in the complex case. Here, only the principal facts underlying the proof of Theorem \ref{MMLE_asyn} are discussed. 
To prove the consistency of $\hat{\alpha}$, we need an extension of the Law of Large Numbers (LLN) to uniform and strong mixing random sequences (see Assumption \ref{assumption_mix}). This result is stated in Theorem 2.3 of \cite{white_nl_reg_dep}. With this suitable generalization of the LLN, the (strong) consistency of the LS estimator can be readily established as shown by Theorem 3.1 in \cite{white_nl_reg_dep}. Regarding the asymptotic normality, a generalization to uniform and strong mixing random sequences of the Central Limit Theorem (CLT) is required. This extension can be found in \cite[Th. 2.4]{white_nl_reg_dep} for the scalar case, and in \cite[Th. 2]{Multivar_CLT_1} for the multivariate case. The asymptotic normality of $\hat{\alpha}$ can be established by a direct application of this version of the CLT as shown in \cite[Th. 3.2]{white_nl_reg_dep}. Finally, the extension of these results to the complex field can be obtained simply by exploiting the natural set isomorphism between $\mathbb{C}$ and $\mathbb{R}^2$ and by using the fact that only circular random sequences are considered.         	
\end{IEEEproof}

\begin{remark}\label{remark2}
	Note that, since $\hat{\alpha}$ is obtained as a linear combination of circular observations ${x_1, \ldots, x_N}$, its second-order statistics are fully characterized by the variance $E_{p_X}\{|\hat{\alpha} - \bar{\alpha}|^2\}$, while its pseudo-variance $E_{p_X}\{(\hat{\alpha} - \bar{\alpha})^2\}$ is nil \cite{Pici}.
\end{remark}

While $A_N$ in \eqref{A_def} is only a function of the known norm of $\mb{v}$, $\bar{B}_N$ in \eqref{B_def} requires to compute the expectation w.r.t. the true, but unknown, disturbance pdf $p_C$. 
It can be shown that, under a uniform (or strong) mixing condition for the disturbance process $\{c_n: \forall n\}$, a consistent estimator of $\bar{B}_N$ is \cite{white_nl_reg_dep}:
\begin{align}\notag\label{B_est}
\hat{B}_N&\equiv\hat{B}_N(\hat{\alpha}) = N^{-1}\sum\limits_{n=1}^N|v_n|^2|\hat{c}_n|^2 \\&+ 2N^{-1}\sum\limits_{m=1}^{l}\sum\limits_{n=m+1}^N\realpart{v_nv^*_{n-m} \hat{c}_n\hat{c}_{n-m}^*}
\end{align}
where $l$ is the so-called \textit{truncation lag} \cite{white_nl_reg_dep},
\be
\hat{c}_n = x_n - \hat{\alpha}v_n, \quad \forall n
\ee
and $\hat{\alpha}$ is given in \eqref{alpha_est}. The estimate $\hat{B}_N$ in \eqref{B_est} can be rewritten in a more compact form as:
\be
\hat{B}_N = N^{-1}\mb{v}^H\widehat{\bs{\Gamma}}_l\mb{v},
\ee 
where
\be\label{C_est}
[\widehat{\bs{\Gamma}}_l]_{i,j} \triangleq \left\lbrace  \begin{array}{cc}
	\hat{c}_i\hat{c}_j^* & j-i \leq l\\
	\hat{c}_i^*\hat{c}_j & i-j \leq l\\
	0 & |i-j| > l
\end{array} \right.
\ee
for $1\leq i,j \leq N$. The consistency of $\hat{B}_N$ is stated in the next theorem (see \cite[Th. 3.5]{white_nl_reg_dep}).
\begin{theorem}\label{cons_hat_B}
	Under Assumption \ref{assumption_mix}, if $l \rightarrow \infty$ as $N \rightarrow \infty$ such that $l=o(N^{1/3})$ then \footnote{Given a real-valued function $f(x)$ and a strictly positive real-valued function $g(x)$, $f(x)=o(g(x))$ if for every positive real number $a$, there exists a real number $x_0$ such that $|f(x)| \leq a g(x),\; \forall x \geq x_0$.}:
	\be
	\hat{B}_N - \bar{B}_N \overset{p}{\underset{N\rightarrow \infty}{\rightarrow}} 0.
	\ee
\end{theorem}
\begin{IEEEproof}
	The proof is given in \cite[Th. 6.20]{White_B}.
\end{IEEEproof}

\smallskip
Theorem \ref{cons_hat_B} provides us with a useful criterion to choose the truncation lag $l$. In fact, to ensure the consistency of $\hat{B}_N$, $l$ has to grow with $N$, but more slowly than $N^{1/3}$.

\subsection{A robust Wald-type test}
Given the asymptotic characterization of the LS estimator presented in Theorem \ref{MMLE_asyn}, the Wald-type statistic can be set up as:
\be
\label{Wald_test}
\Lambda_\mathsf{RW}(\mb{x}) = \frac{2N|\hat{\alpha}|^2}{A_N^{-2}\hat{B}_N} = \frac{2|\mb{v}^H\mb{x}|^2}{ \mb{v}^H\widehat{\bs{\Gamma}}_l\mb{v} },
\ee
where the entries of $\widehat{\mb{\Gamma}}_l$ are given in \eqref{C_est}. The similarity between the proposed $\Lambda_\mathsf{RW}$ statistic and the Adaptive Matched Filter (AMF) proposed in \cite{CFAR_AMF} is evident. However, the following comments are in order. 

First, in \cite{CFAR_AMF} a set of \textit{homogeneous secondary snapshots}, i.e. a set of \virg{signal-free} data collected from radar resolution cells adjacent to the cell under test or at different time instants, is used to estimate the disturbance covariance matrix. This approach, however, requires that the disturbance statistics remain constant over all the considered resolution cells or time instants. Unlike the AMF, the decision statistic $\Lambda_\mathsf{RW}$ in \eqref{Wald_test} does not need any secondary data since it is able to extract all the required information from the \textit{single snapshot} collected in the cell under test. Secondly, the AMF in \cite{CFAR_AMF} is derived under the Gaussian assumption for the disturbance vector $\mb{c}$. Conversely, $\Lambda_\mathsf{RW}$ in \eqref{Wald_test} can handle all the disturbance distributions that satisfy Assumption \ref{assumption_mix}, including the Gaussian one. Third, as shown in Theorem \ref{Theo_MW} below, $\Lambda_\mathsf{RW}$ in \eqref{Wald_test} is shown to have the Constant False Alarm Rate (CFAR) property as $N\to \infty$ for all the disturbance distributions satisfying Assumption \ref{assumption_mix} and without the need of any secondary data. This is a great advantage w.r.t. the AMF that is CFAR only if the disturbance is Gaussian-distributed and a set of homogeneous secondary data is available.

The asymptotic property of $\Lambda_\mathsf{RW}(\mb{x})$ can be stated as follows. 
\begin{theorem}
	\label{Theo_MW}
	If Assumption \ref{assumption_mix} hold true, then	\begin{align}
	\Lambda_{\mathsf{RW}}(\mb{x}|H_0) \underset{N\rightarrow \infty}{\sim} \chi_2^2(0)\label{MW_H0},\\
	\Lambda_{\mathsf{RW}}(\mb{x}|H_1) \underset{N\rightarrow \infty}{\sim} \chi_2^2\tonde{\varsigma}\label{MW_H1},
	\end{align}
	where $\varsigma \triangleq 2|\bar{\alpha}|^2\frac{\norm{\mb{v}}^4}{\mb{v}^H\bs{\Gamma}\mb{v}}$.
\end{theorem} 
\begin{IEEEproof}
	The proof follows from Theorem \ref{MMLE_asyn} and a known result about circular Gaussian random variables \cite{Complex_Normal}. In particular, if $a \sim \mathcal{CN}(\mu_a,\sigma_a^2)$, then $2|a-\mu_a|^2/\sigma_a^2\sim \chi_2^2(0)$. Details are given in Appendix B.
\end{IEEEproof}
\smallskip
The above theorem shows that the pdfs of \eqref{Wald_test} under $H_0$ and $H_1$ converge to $\chi$-squared pdfs with $2$ degrees of freedom when the number of virtual spatial antenna channels $N = M_T M_R$ goes to infinity. This means that \eqref{int_eq_1} is asymptotically satisfied by $\bar{\lambda} = -2\ln \overline{P_{FA}}$. This is valid for any pre-assigned $\overline{P_{FA}}$ and, more importantly, for any disturbance process $\{c_n: \forall n\}$ satisfying Assumption \ref{assumption_mix}. In other words, we could say that $\Lambda_{\mathsf{RW}}(\mb{x})$ achieves the CFAR property w.r.t. all the disturbance distributions satisfying Assumption \ref{assumption_mix} when a sufficiently large number of transmitting and receiving antennas is used. Numerical results will be used to show that such \emph{Massive MIMO regime} is achieved for a feasible large number of antennas. Some considerations on the asymptotic distribution of $\Lambda_{\mathsf{RW}}(\mb{x})$ under $H_1$ can also be done. In particular, in order to make explicit the dependence of $\varsigma$ from $\mb{a}_T(\bar{\bs{\phi}})$, $\mb{a}_R(\bar{\bs{\phi}})$ and $\mb{W}$, one can substitute the definition of $\mb{v}$ given in \eqref{v_vect} into $\varsigma$ to obtain:
\begin{align}\label{delta_1}
\varsigma = \frac{2|\bar{\alpha}|^2M_R^2\|(\mb{W}\mb{S})^T\mb{a}_T(\bar{\bs{\phi}})\|^4}{\mathrm{tr}\tonde{\bs{\Gamma}\quadre{ (\mb{W}\mb{S})^T\mb{a}_T(\bar{\bs{\phi}})\mb{a}_T^H(\bar{\bs{\phi}})(\mb{W}\mb{S})^* \otimes \mb{a}_R(\bar{\bs{\phi}})\mb{a}_R^H(\bar{\bs{\phi}})}}}.
\end{align}

\begin{remark} \label{rem_W}
Further manipulations to the expression of $\varsigma$ in \eqref{delta_1} are allowed if the model in \eqref{C_int} is adopted for the covariance matrix $\bs{\Gamma}$. Specifically, by substituting \eqref{C_int} in \eqref{delta_1}, we get:
	\be\label{delta_2}
	\!\!\varsigma = \frac{2|\bar{\alpha}|^2M_R\|(\mb{W}\mb{S})^T\mb{a}_T(\bar{\bs{\phi}})\|^2}{\iint_{0}^{T}  \norm{\mb{s}_o(t-\bar{l}\Delta t)}^2 \mathrm{tr}\quadre{\bs{\Sigma}(t-\tau)}  e^{-j\bar{k} \Delta\omega (t-\tau)}dtd\tau},
	\ee
	where $\bar{l}$ and $\bar{\omega}$ defines the range-Doppler gate under test. Moreover, if $\mb{s}_o(t)$ is a vector of perfectly orthonormal waveforms, i.e. $\mb{S}=\mb{I}_{M_T}$, and if $\bs{\Sigma}(t-\tau) = \sigma^2\mb{I}_{M_R}\delta(t-\tau)$, \eqref{delta_2} can be further simplified as:
	\be\label{delta_3}
	\varsigma = \frac{2|\bar{\alpha}|^2 P(\bar{\bs{\phi}})^2 }{\sigma^2},
	\ee
	where $P(\bs{\phi}) \triangleq \mb{a}_T^H(\bs{\phi})\mb{W}^*\mb{W}^T\mb{a}_T(\bs{\phi})$ is the beam pattern of the transmitting array as a function of the position vector $\bs{\phi}$. The same result can be found in \cite{MIMO_TAB}, \textcolor{blue}{where a LR test, perfectly matched to white Gaussian disturbance, was exploited as a detector.}
\end{remark}
\smallskip

\textcolor{blue}{The expression in \eqref{delta_3} suggests us an interesting fact. Under the specific and simplistic scenario of perfectly orthonormal waveforms and (spatial and temporal) white Gaussian disturbance, a perfectly matched LR-based detector has the same (asymptotic) detection performance of the robust Wald-type test proposed here in \eqref{Wald_test}. In other words, under the simplistic scenario mentioned above, $\Lambda_\mathsf{RW}$ in \eqref{Wald_test} does not present any loss in detection w.r.t. the \virg{optimal} LR test even if $\Lambda_\mathsf{RW}$ does not require any a priori knowledge about the Gaussianity of the disturbance, while the LR test does. It is also worth mentioning that, in a more involved and realistic scenario, an LR-based test is practically unfeasible due to the lack of a priori information on the functional form of the disturbance pdf. Moreover, even if such disturbance pdf were available, the derivation of LR statistics would likely be met with the impossibility to derive its closed form expression. On the contrary, the closed form expression and the asymptotic detection performance of $\Lambda_\mathsf{RW}$ in \eqref{Wald_test} remain unchanged under any known or unknown disturbance process satisfying Assumption \ref{assumption_mix}.}   

The following corollary is found.
\begin{corollary}
	\label{cor}
	If Assumption \ref{assumption_mix} holds true, then the detection probability of \eqref{Wald_test} is such that
	\be
	\label{closed_form_ROC}
	P_D(\lambda) \to_{N\to\infty} Q_1\tonde{\sqrt{\varsigma},\sqrt{\lambda}},
	\ee 
	where $Q_1(\cdot,\cdot)$ is the Marcum Q function of order 1 \cite{non_cenrtal_CDF} and $\varsigma$ is still given by $\varsigma = 2|\bar{\alpha}|^2\frac{\norm{\mb{v}}^4}{\mb{v}^H\bs{\Gamma}\mb{v}}$.
\end{corollary}
\begin{IEEEproof}
	By definition 
	\be
	P_D(\lambda)\triangleq \Pr \graffe{\Lambda(\mb{X}) \geq \lambda | H_1}=\int_{\lambda}^{\infty} p_{\Lambda|H_1}(a|H_1)da. 
	\ee
	Then, \eqref{closed_form_ROC} follows directly from \eqref{MW_H1} and the properties of the non-central $\chi^2$ distribution \cite{non_cenrtal_CDF}.
\end{IEEEproof}

\smallskip
Since $Q_1(\cdot,\cdot)$ is monotonic in its first argument, Corollary \ref{cor} states that the $P_D$ of the RW test in \eqref{Wald_test} goes to 1 as $N\to\infty$. Moreover, it shows that $P_D$ depends on the true, but unknown, covariance matrix $\mb{\Gamma}$ of the disturbance vector $\mb{c}$, the radar geometry, and the waveform matrix $\mb{W}$ (through the vector $\mb{v}$).

\vspace{-0.2cm}
\begin{figure}[t!]
	\centering
	\includegraphics[width=0.9\columnwidth]{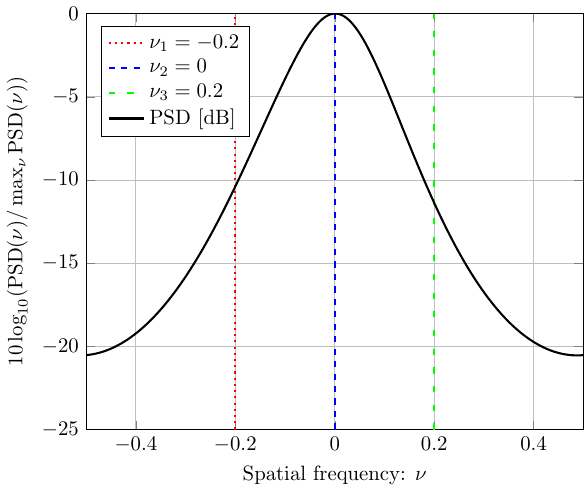}
	\caption{PSD of the $\mathsf{AR}(3)$ in Scenario 1.}\vspace{-0.3cm}
	\label{fig:Fig2}
\end{figure}

\section{Numerical Analysis}

Numerical results are now used to validate the theoretical findings on the asymptotic properties of $\Lambda_{\mathsf{RW}}$ as stated in Theorem \ref{Theo_MW}. We consider a uniform linear array at the transmitter and receiver, and a single target located in the far-field. We assume that $\mb{W} = \mb{I}_{M_T}$ and the transmitted waveforms are orthogonal, i.e., $\mb{S} = \mb{I}_{M_T}$. Following \cite{Stoica_ident}, we choose the radar geometry that maximizes the parameter identifiability. This is achieved by using a receiving array characterized by $M_R$ antenna elements with an inter-element spacing of $d$ and a transmitting array whose $M_T$ elements are spaced by $M_R d$. This implies that 
\be\label{a_r_V}
\mb{a}_R(\bs{\phi}) = [1,e^{\mathsf{j}2\pi \nu},\ldots,e^{\mathsf{j}2\pi(M_R-1)\nu}]^T,
\ee
\be\label{a_t_V}
\mb{a}_T(\bs{\phi}) = [1,e^{\mathsf{j}2\pi M_R \nu},\ldots,e^{\mathsf{j}2\pi(M_T-1)M_R\nu}]^T,
\ee
where the \textit{spatial frequency} $\nu$ is
\be
\nu \triangleq \frac{f_0d}{c}\sin(g(\bs{\phi})),
\ee
where $f_0$ is the carrier frequency of the transmitted signal, $c$ is the speed of light and $g(\cdot)$ is a known function of the position vector $\bs{\phi}$. By substituting \eqref{a_r_V} and \eqref{a_t_V} into \eqref{v_vect}, the vector $\mb{v}(\bs{\phi})\in \mathbb{C}^{N}$ can be expressed as, for $i = 1,\ldots,N= M_RM_T$
\be\label{v_s2}
[\mb{v}(\bs{\phi})]_i= e^{j2\pi(i-1)\nu}
\ee
which represents the steering vector of an equivalent phased-array with $N$ elements \cite{Stoica_ident}.

Numerical results are obtained by averaging over $10^6$ Monte Carlo simulations. Moreover, the truncation lag in Theorem \ref{cons_hat_B} is chosen as $l = \lfloor N^{1/4}\rfloor$.

\begin{figure}[t!]
	\centering
	\includegraphics[width=0.9\columnwidth]{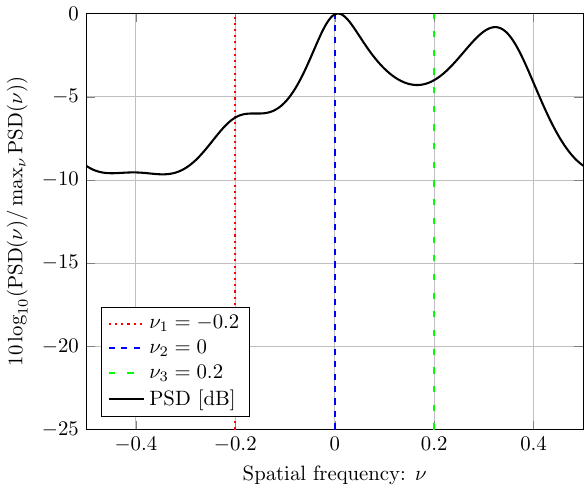}
	\caption{PSD of the $\mathsf{AR}(6)$ in Scenario 2.}\vspace{-0.3cm}
	\label{fig:Fig3}
\end{figure}



\subsection{Disturbance models}
Two different models are considered for the disturbance. In Scenario 1, the disturbance vector $\mb{c}$ is generated according to an underlying circular, SOS $\mathsf{AR}(p)$
\be
\label{cn_2}
c_n = \sum\nolimits_{i=1}^{p}\bar{\rho}_i c_{n-i} + w_n, \quad n \in (-\infty,\infty),
\ee
with $p=3$, driven by i.i.d., $t$-distributed innovations $w_n$ whose pdf $p_w$ is \cite{TSP_MCRB}:
\be
p_{w}(w_n) = \frac{\lambda}{\sigma_w^2 \pi} \tonde{\frac{\lambda}{\eta}}^{\lambda}\tonde{\frac{\lambda}{\eta} + \frac{|w_n|^2}{\sigma_w^2}}^{-(\lambda+1)},
\ee
where $\lambda \in (1,\infty)$ and $\eta = \lambda/(\sigma_w^2(\lambda-1))$ are the shape and scale parameters. Specifically, $\lambda$ controls the tails of $p_{w}$. If $\lambda$ is close to 1, then $p_{w}$ is heavy-tailed and highly non-Gaussian. On the other hand, if $\lambda \rightarrow \infty$, then $p_{w}$ collapses to the Gaussian distribution. In our simulations, we set $\lambda = 2$ and $\sigma_w^2 = 1$. The $\mathsf{AR}(3)$ coefficient vector is $\bar{\bs{\rho}} = [ 0.5e^{\mathsf{j}2\pi0}, 0.3e^{-\mathsf{j}2\pi0.1}, 0.4e^{\mathsf{j}2\pi0.01}]^T$. The normalized PSD can be expressed as
can be expressed as:
\be\label{PSD_nu_2}
S(\nu) \triangleq \sigma_w^2 \left| 1- \sum\nolimits_{n=1}^{p}\bar{\rho}_n e^{-\mathsf{j}2\pi n \nu}  \right|^{-2}, \quad p=3, 
\ee
and is shown in Fig. \ref{fig:Fig2}. As seen, Scenario 1 is characterized by a disturbance whose power is mostly concentrated around the spatial frequency $\nu = 0$. 

To prove the robustness of the proposed $\Lambda_{\mathsf{RW}}$ w.r.t. more general disturbance models, in Scenario 2 we increase the order of the AR process generating the disturbance vector $\mb{c}$. Particularly, we consider a circular, SOS $\mathsf{AR}(6)$ driven (as before) by i.i.d., $t$-distributed innovations $w_n$ and characterized by the following coefficient vector $\bar{\bs{\rho}} = [ 0.5e^{-\mathsf{j}2\pi0.4}, 0.6e^{-\mathsf{j}2\pi0.2}, 0.7e^{\mathsf{j}2\pi0}, 0.4e^{\mathsf{j}2\pi0.1}, 0.5e^{\mathsf{j}2\pi0.3},$ $0.6e^{\mathsf{j}2\pi0.35} ]^T$. The normalized PSD is reported in Fig. \ref{fig:Fig3} and shows that, differently from Scenario 1, the disturbance power is spread over the whole range of $\nu$. Moreover, it presents more than a single peak. 


%
\begin{figure}[t!]
	\centering
	\includegraphics[width=0.9\columnwidth]{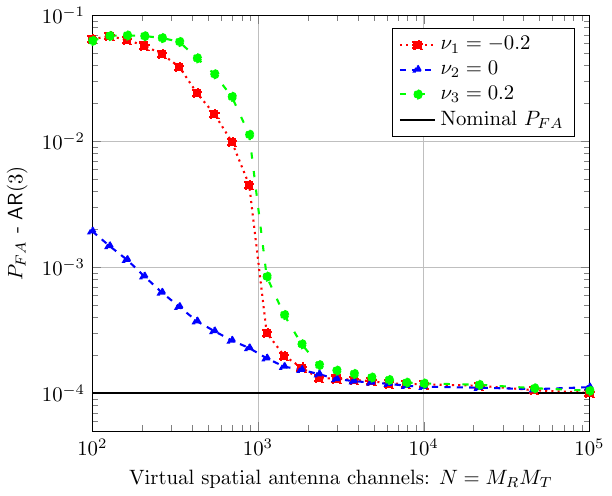}
	\caption{Estimated $P_{FA}$ as a function of the virtual spatial antenna channels $N$ in Scenario 1.}\vspace{-0.3cm}
	\label{fig:Fig4}
\end{figure}

In both scenarios, the disturbance process is normalized in order to have $\sigma^2 = r_C[0]=1$. Under hypothesis $H_1$, the signal-to-noise ratio is simply defined as $\mathrm{SNR}\triangleq 10 \log_{10} (|\bar{\alpha}|^2/\sigma^2)$. 

\subsection{Performance Analysis}
\begin{figure}[t!]
	\centering
	\includegraphics[width=0.9\columnwidth]{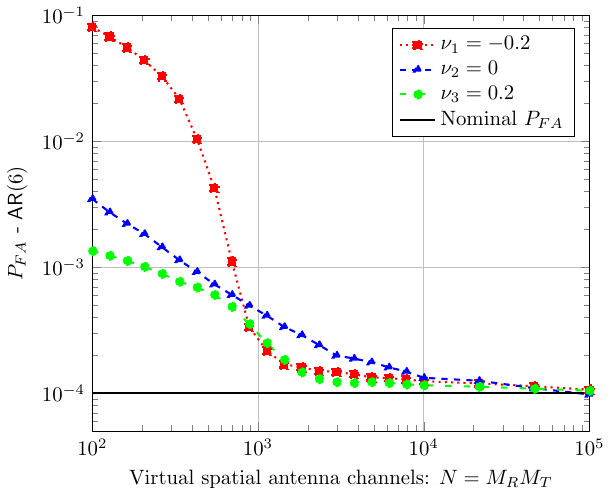}
	\caption{Estimated $P_{FA}$ as a function of the virtual spatial antenna channels $N$ in Scenario 2.}\vspace{-0.3cm}
	\label{fig:Fig5}
\end{figure}

Since the disturbance PSD in Figs. \ref{fig:Fig2} and \ref{fig:Fig3} is not constant w.r.t the spatial frequency $\nu$, the performance of $\Lambda_{\mathsf{RW}}(\mb{x})$ will be evaluated for three different values of $\nu$ corresponding to different disturbance power density levels (i.e. three different target DOAs): $\nu_1 = -0.2$, $\nu_1 = 0$ and $\nu_1 = 0.2$. Figs. \ref{fig:Fig4} and \ref{fig:Fig5} plot the $P_{FA}$ of the proposed robust Wald test in \eqref{Wald_test} for the three considered values of $\nu$. As we can see, in both scenarios $\Lambda_{{\mathsf{RW}}}$ in \eqref{Wald_test} achieves the nominal value of $10^{-4}$ for $N \geq 10^4$. Moreover, in this Massive MIMO regime, i.e. for $N \geq 10^4$, the progress of the $P_{FA}$ curves for different scenarios and for different values of spatial frequency are almost identical. This provides a numerical validation of the theoretical result provided in \eqref{MW_H0} of Theorem \ref{Theo_MW}. 

\begin{figure}[t!]
	\centering
	\includegraphics[width=0.9\columnwidth]{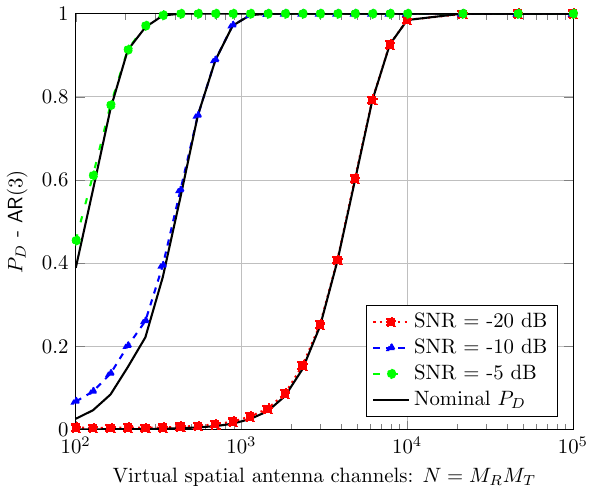}
	\caption{Estimated and nominal $P_{D}$ as a function of the virtual spatial antenna channels $N$ in Scenario 1 for different SNR values, spatial frequency $\nu = 0$, and nominal $P_{FA}=10^{-4}$.}\vspace{-0.3cm}
	\label{fig:Fig6}
\end{figure}

Fig. \ref{fig:Fig6} considers Scenario 1 and illustrates the estimated and the closed-form expression of $P_{D}$ given in Corollary \ref{cor} for three distinct SNR values. Particularly, we assume $\mathrm{SNR} = -20, -10$ and $-5$ dB. As seen, the $P_{D}$ tends to $1$ as the number of virtual spatial antenna channels $N$ increases. Specifically, for $\mathrm{SNR} \ge -20$ dB the $P_{D}$ approaches $1$ for $N \geq 10^4$. From Fig. \ref{fig:Fig6}, it is also immediate to verify that the $P_{D}$ estimated through Monte-Carlo runs is in perfect agreement with the theoretical one provided in Corollary \ref{cor}. We conclude by noticing that similar numerical results have been obtained for the CG disturbance model discussed in Sub-section \ref{c_model}. Since they are perfectly in line with the numerical analysis reported above, we decided to not include them here due to lack of space.

\section{Concluding remarks and discussions}

The detection problem in co-located MIMO radar systems was analysed in this paper. A robust Wald-type detector was proposed and its asymptotic performance investigated when the virtual spatial antenna channels $N=M_TM_R$ goes to infinity. \textcolor{blue}{Specifically, the CFAR property of the proposed detector for the asymptotic regime $N\to \infty$ and the wide family of disturbance processes satisfying Assumption \ref{assumption_mix} was mathematically proved and validated through numerical simulations.}  
{\color{blue}The purpose of analysing the asymptotic performance when $N\to \infty$ is not that we advocate the deployment of radars with a nearly infinite number of virtual antennas. The importance of asymptotics is instead what it tells us about practical systems with a finite number of antennas. Indeed, our main results imply that we can always satisfy performance requirements by deploying sufficiently many virtual antennas $N$, without any a priori knowledge of the disturbance statistics. Our numerical results showed that a pre-assigned value of $\overline{P_{FA}}=10^{-4}$ can be achieved with $N=M_TM_R\ge 10^4$ with non-Gaussian, stable autoregressive disturbance models of order $p = 3$ and $6$. This defines the so-called Massive MIMO regime of the radar.} 

In Massive MIMO communications, linear combining, and precoding schemes can entirely eliminate the interference as the number of antennas grows unboundedly even with imperfect knowledge of propagation channels. We showed that a large-scale MIMO radar yields a target detector, which is robust to the unknown disturbance statistics. We foresee that further breakthroughs can be obtained by extending the Massive MIMO concept to other radar problems \cite{Bjornson2019a}. Clearly, by using very large arrays for waveform design, one can radically improve the spatial diversity gain and spatial resolution for target detection, parameter estimation, and interference rejection. However, the lesson learned from the last decade of research in communications is that Massive MIMO is not merely a system with many antennas but rather a paradigm shift with regards to the modelling, operation, theory, and implementation of MIMO systems. \textcolor{blue}{Our vision is that this paradigm can be applied also to radars, and open new research directions.}


\appendices

\section{Proof of Theorem \ref{MMLE_asyn}}
In this Appendix, the main ideas behind the proof of Theorem \ref{MMLE_asyn} are discussed. Specifically, we show the $\sqrt{N}$-\textit{consistency} and the \textit{asymptotic normality} of the mismatched LS estimator $\hat{\alpha}$ in \eqref{alpha_est} under Assumption \ref{assumption_mix}.
\subsection{Consistency}  
Let us recall here the explicit expression of the LS objective function in \eqref{obj_fun}: $
G_N(\mb{x},\alpha) \triangleq \sum\nolimits_{n=1}^{N} |x_n - \alpha v_n|^2$.
Moreover, let $\bar{G}_N(\alpha)$ be the following measurable and continuous function: 
\be\label{Q_bar}
\bar{G}_N(\alpha) \triangleq \sum\nolimits_{n=1}^NE_{p_{X}}\graffe{|x_n -  \alpha v_n|^2}.
\ee
Then, under Assumption \ref{assumption_mix}, from \cite[Th. 2.3]{white_nl_reg_dep} and \cite[Th. 1]{ULLN}, we have that
\be\label{conv_q}
\underset{\alpha \in \Omega}{\sup}\frac{1}{N}\left|{G_N(\mb{x},\alpha) -  \bar{G}_N(\alpha) }\right| \overset{a.s.}{\underset{N\rightarrow \infty}{\rightarrow}} 0,
\ee 
where $\Omega$ is a compact subset of $\mathbb{C}(\equiv \mathbb{R}^2)$. For the LS estimator $\hat{\alpha}$ in \eqref{alpha_est}, the following convergence property holds.
\begin{theorem}
	If $\bar{G}_N(\alpha)$ has a unique minimum at $\alpha_{0,N}\in \mathbb{C}$, then from \eqref{conv_q}:
	\be
	\frac{1}{N}\left|{G_N(\mb{x},\hat{\alpha}) - \bar{G}_N(\alpha_{0,N})}\right| \overset{a.s.}{\underset{N\rightarrow \infty}{\rightarrow}} 0 
	\ee
	and
	\be
	\hat{\alpha} - \alpha_{0,N} \overset{a.s.}{\underset{N\rightarrow \infty}{\rightarrow}} 0 
	\ee
	where $\alpha_{0,N} \triangleq \underset{\alpha \in \mathbb{C}}{\mathrm{argmin}}\; \bar{G}_N(\alpha)$.
\end{theorem}
\begin{IEEEproof}
	See the proof of Theorem 2.4 in \cite{white_theo}.
\end{IEEEproof}
Note that $\alpha_{0,N}$ represents the counterpart of the \textit{pseudo-true parameter} defined in \cite{white} for the i.i.d. data case.
The consistency of $\hat{\alpha}$ can finally be established by showing that the pseudo-true parameter $\alpha_{0,N}$ is equal to the true one $\bar{\alpha}$. To this end, from \eqref{Q_bar} we obtain:
\begin{align}\notag
\bar{G}_N(\alpha) &= \sum_{n=1}^{N}E_{p_{X}}\graffe{|x_n -  \alpha v_n|^2} \\\notag
	&= \sigma_c^2 -|\bar{\alpha} - \alpha|^2\norm{\mb{v}}^2 \times\\
	& 2\realpart{(\bar{\alpha} - \alpha)\textstyle\sum\limits_{n=1}^{N}E_{p_{X}}\graffe{x_n -  \bar{\alpha} v_n}v_n}.
\end{align}
By definition of data process in \eqref{HT_base_x}, $E_{p_{X}}\graffe{x_n -  \bar{\alpha} v_n}=0,\; \forall n$, then the minimum of $\bar{Q}_N(\alpha)$ is attained at $\alpha_{0,N} = \bar{\alpha}$. This proves the consistency of $\hat{\alpha}$ under Assumption \ref{assumption_mix}.

\subsection{Asymptotic normality}
The proof for the asymptotic normality of the LS estimator $\hat{\alpha}$ mimics the one provided in standard statistical textbooks (see e.g. \cite[Ch. 9]{Cox}) for the Maximum Likelihood estimator. Let us start by taking the Taylor's expansion of the real-valued objective function $G_N(\mb{x},\alpha)$ around the complex parameter $\bar{\alpha}$ \cite[Th. 3.3]{Complex_ollila}:
\be
\begin{split}
	\!\!\!\!G_N&(\mb{x},\alpha) \backsimeq  G_N(\mb{x},\bar{\alpha})+ (\alpha-\bar{\alpha})\left. \parder{G_N(\mb{x},\alpha)}{\alpha}\right|_{\alpha = \bar{\alpha}} \!\!\!\!\!\!\!\!+ \\ +& (\alpha-\bar{\alpha})^*\left. \parder{G_N(\mb{x},\alpha)}{\alpha^*}\right|_{\alpha = \bar{\alpha}}
	\!\!\!\!\!\!\!\!+ |\alpha-\bar{\alpha}|^2 \left. \frac{\partial^2G_N(\mb{x},\alpha)}{\partial\alpha\partial\alpha^*}\right|_{\alpha = \bar{\alpha}}\!\!\!\!,
\end{split}
\ee
where we used the fact that:
\be
\frac{\partial^2G_N(\mb{x},\alpha)}{\partial\alpha^2} = \frac{\partial^2G_N(\mb{x},\alpha)}{\partial\alpha^{*2}} = 0, \; \forall \alpha.
\ee
By using the Mean Value Theorem \cite[Ch. 9]{Cox}, for some $\tilde{\alpha}$ such that $|\tilde{\alpha}-\bar{\alpha}|<|\hat{\alpha}-\bar{\alpha}|$ we have: 
\be
\left. \parder{G_N(\mb{x},\alpha)}{\alpha^*}\right|_{\alpha = \bar{\alpha}} + (\hat{\alpha}-\bar{\alpha}) \left.\frac{\partial^2G_N(\mb{x},\alpha)}{\partial\alpha\partial\alpha^*}\right|_{\alpha = \tilde{\alpha}} = 0,
\ee 
where $\hat{\alpha}$ is defined in \eqref{MMLE}. Consequently, we obtain
\be
\label{MV_exp}
\hat{\alpha} - \bar{\alpha} = - \tonde{\left.\frac{\partial^2G_N(\mb{x},\alpha)}{\partial\alpha\partial\alpha^*} \right|_{\alpha = \tilde{\alpha}}}^{-1}\left.\parder{G_N(\mb{x},\alpha)}{\alpha^*}\right|_{\alpha = \bar{\alpha}}.
\ee
For simplicity, we define
\be
A_N(\mb{x},\alpha) \triangleq \frac{1}{N}\frac{\partial^2G_N(\mb{x},\alpha)}{\partial\alpha\partial\alpha^*},
\ee
\be
s(\mb{x},\alpha) \triangleq \parder{G_N(\mb{x},\alpha)}{\alpha^*},
\ee
and rewrite \eqref{MV_exp} as
\be
\label{MV_exp_2}
\sqrt{N}(\hat{\alpha} - \bar{\alpha}) = -A_N(\mb{x},\tilde{\alpha})^{-1}\tonde{\frac{1}{\sqrt{N}}s(\mb{x},\bar{\alpha})}.
\ee
Through direct calculation, it easily follows that
\begin{align}
A_N(\mb{x},\alpha) &= N^{-1} \sum\limits_{n=1}^N|v_n|^2 = N^{-1}\norm{\mb{v}}^2 \triangleq A_N\label{A_N},\\\label{score}
s(\mb{x},\bar{\alpha}) &= -\sum\limits_{n=1}^Nv_n^*\tonde{x_n-\bar{\alpha}v_n} = - \sum\limits_{n=1}^Nv_n^*c_n,
\end{align}
where $c_n = x_n - \bar{\alpha}v_n$,$\forall n$. By substituting \eqref{A_N} and \eqref{score} into \eqref{MV_exp_2}, we get:
\be
\sqrt{N}(\hat{\alpha} - \bar{\alpha}) = \frac{N}{\norm{\mb{v}}^2}\tonde{\frac{1}{\sqrt{N}}\sum\limits_{n=1}^Nv_n^*c_n}. 
\ee
The asymptotic normality of $\hat{\alpha}$ follows from the application of the Central Limit Theorem established for (strong) mixing processes in \cite[Th. 2.4]{white_nl_reg_dep} for the real scalar case and in \cite{Multivar_CLT_1} for the real multivariate case. In particular, let us define
\begin{align}\notag
\!\!\!\!\!\!\bar{B}_N &\equiv B_N(\bar{\alpha}) \triangleq E_{p_X}\left\{\frac{1}{\sqrt N}s(\mb{x},\bar{\alpha})\frac{1}{\sqrt N}s^*(\mb{x},\bar{\alpha})\right\} \\\label{B_N_theo}
&= \frac{1}{N}\sum\limits_{n=1}^N\sum\limits_{m=1}^Nv_n^*v_m E_{p_X}\{c_nc_m^*\}.\!\!\!\!\!\!
\end{align}
Under Assumption \ref{assumption_mix}, by exploiting \cite[Th.3.2]{white_nl_reg_dep} and recalling that $\{c_n:\forall n\}$ is a circular process, \eqref{asym_norm} follows.



\section{Asymptotic distribution of $\Lambda_{\mathsf{RW}}$}
To establish the asymptotic distribution of $\Lambda_{\mathsf{RW}}$, we follow the standard procedure discussed in \cite[Ch. 9]{Cox}. 

\subsection{Asymptotic distribution of $\Lambda_{\mathsf{RW}}$ under $H_0$}
We start by defining
\be\label{bar_alpha_H0}
\bar{\alpha}_{H_0} \triangleq 0
\ee
as the true parameter under the null hypothesis. Under $H_0$, from Theorem 1, we have that:
\be
\hat{\alpha}\overset{a.s.}{\underset{N\rightarrow \infty}{\rightarrow}} \bar{\alpha}_{H_0},
\ee
\be
\label{norm_alpha}
\sqrt{N}A_N\quadre{B_N(\bar{\alpha}_{H_0})}^{-1/2}\hat{\alpha} \underset{N\rightarrow \infty}{\sim} \mathcal{CN}(0,1).
\ee
The inverse operator and the principal square root are both continuous operators on $\mathbb{R}^+$, then their composition is continuous on $\mathbb{R}^+$. Then, from \eqref{Wald_test} and by using the Continuous Mapping Theorem \cite[Theo. 2.7]{Billi}, it follows that:
\begin{align}
\notag
\quadre{\hat{B}_N(\hat{\alpha})}^{-1/2}&-\quadre{B_N(\bar{\alpha}_{H_0})}^{-1/2} \\
& \equiv\hat{B}_N^{-1/2} - \quadre{B_N(0)}^{-1/2} \overset{p}{\underset{N\rightarrow \infty}{\rightarrow}} 0.\label{conv_C_est_root}
\end{align}
Let us rewrite the test in \eqref{Wald_test} as:
\be
\label{WT_2}
\Lambda_{\mathsf{RW}}(\mb{x}) = 2 \tonde{\sqrt{N}A_N\hat{B}_N^{-1/2}\hat{\alpha}}^*\tonde{\sqrt{N}A_N\hat{B}_N^{-1/2}\hat{\alpha}}.
\ee
From \eqref{norm_alpha} and \eqref{conv_C_est_root}, by a direct application of the Slutsky's Theorem and of the properties of the complex Gaussian distribution \cite{Complex_Normal}, we immediately obtain that:
\begin{align}\notag
	\Lambda_{\mathsf{RW}}(\mb{x}|H_0) =& 2{\underbrace{\tonde{\sqrt{N}A_N\hat{B}_N^{-1/2}\hat{\alpha}}}_{\underset{N\rightarrow \infty}{\sim} \mathcal{CN}\tonde{0,1}}}^*\times\\\label{asy_H0_MWT}
	&\underbrace{\tonde{\sqrt{N}A_N\hat{B}_N^{-1/2}\hat{\alpha}}}_{\underset{N\rightarrow \infty}{\sim} \mathcal{CN}\tonde{0,1}} \underset{N\rightarrow \infty}{\sim} \chi_2^2(0),
\end{align} 
where $\chi_2^2(0)$ indicates a central $\chi$-squared random variable with two degrees of freedom.

\subsection{Asymptotic distribution of $\Lambda_{\mathsf{RW}}$ under local alternatives}

Following \cite[Ch. 9]{Cox}, we suppose that the alternative to $H_0$ is of the form:
\be
\label{alter}
H_1: \; \alpha = d/\sqrt{N}, \quad d \in \mathbb{C}.
\ee 
Note that, as stated in \cite[Sec. 9.3]{Cox}, this choice is made only for mathematical purposes, and has no direct physical significance. Specifically, \eqref{alter} allows us to approximate the \textit{power} of the test (or, in radar terminology, the probability of detection) \textit{locally}, i.e. in the neighbourhood of the null hypothesis in \eqref{bar_alpha_H0}. By defining $\bar{\alpha}_{H_1}^{(N)} \triangleq d/\sqrt{N}$ as the true parameter under local alternatives, we clearly have
\be
\label{lim_theta}
\lim_{N\rightarrow\infty}\bar{\alpha}_{H_1}^{(N)} = \bar{\alpha}_{H_0}.
\ee
If $B_N(\bar{\alpha})$ in \eqref{B_N_theo} is continuous in a neighbourhood of $\bar{\alpha}$, then from \eqref{lim_theta} and \eqref{bar_alpha_H0} it follows that
\be
\label{lim_C}
\lim_{N\rightarrow\infty}B_N(\bar{\alpha}_{H_1}^{(N)}) = B_N(0).
\ee
Under $H_1$ in \eqref{alter}, from Theorem 1 we have
\be
\hat{\alpha} - \bar{\alpha}_{H_1}^{(N)} \overset{a.s.}{\underset{N\rightarrow \infty}{\rightarrow}} 0,
\ee
\be
\label{norm_alpha_H1}
\sqrt{N}A_N\quadre{B_N\tonde{\bar{\alpha}_{H_1}^{(N)}}}^{-1/2}(\hat{\alpha} - d/\sqrt{N}) \underset{N\rightarrow \infty}{\sim} \mathcal{CN}(0,1).
\ee
As before, by using the Continuous Mapping Theorem \cite[Th. 2.7]{Billi} and the limiting results in \eqref{lim_C}, we have
\begin{align}
\notag
\quadre{\hat{B}_N(\hat{\alpha})}^{-1/2}&-\quadre{B_N\tonde{\bar{\alpha}_{H_1}^{(N)}}}^{-1/2} \\
& \equiv\hat{B}_N^{-1/2} - \quadre{B_N(0)}^{-1/2} \overset{p}{\underset{N\rightarrow \infty}{\rightarrow}} 0.\label{conv_C_est_root_H1}
\end{align}
By recasting the test in \eqref{Wald_test} as in \eqref{WT_2}, from \eqref{norm_alpha_H1} and \eqref{conv_C_est_root_H1}, by a direct applications of the Slutsky's Theorem, we immediately obtain that:
\begin{align}\notag
\Lambda_{\mathsf{RW}}(\mb{x}|H_1) &=2 {\underbrace{\tonde{\sqrt{N}A_N\hat{B}_N^{-1/2}\hat{\alpha}}^*}_{\underset{N\rightarrow \infty}{\sim} \mathcal{CN}\tonde{A_N\quadre{B_N\tonde{0}}^{-1/2}d,1}}}\times\\\notag
&\underbrace{\tonde{\sqrt{N}A_N\hat{B}_N^{-1/2}\hat{\alpha}}}_{\underset{N\rightarrow \infty}{\sim} \mathcal{CN}\tonde{A_N\quadre{B_N\tonde{0}}^{-1/2}d,1}}\\\label{asy_H1_MWT}
& \underset{N\rightarrow \infty}{\sim} \chi_2^2\tonde{2|d^2|A_N^2\quadre{B_N(0)}^{-1}}.
\end{align}
Note that, from \eqref{B_N_theo} we have $B_N(0) = N^{-1}\mb{v}^H\bs{\Gamma}\mb{v}$. This result can be used to approximate the power of the test, i.e. the $P_D$. By setting $d \equiv \sqrt{N}\bar{\alpha}$, we have:
\be
\Lambda_{\mathsf{RW}}(\mb{x}|H_1) \underset{N\rightarrow \infty}{\sim} \chi_2^2\tonde{\frac{2|\bar{\alpha}|^2\norm{\mb{v}}^4}{\mb{v}^H\bs{\Gamma}\mb{v}}}.
\ee
By using the properties of the non-central $\chi$-squared distribution with $2$ degrees of freedom \cite{non_cenrtal_CDF}, a closed form expression of the asymptotic probability of detection can be eventually expressed as:
\be
P_D(\lambda) = Q_1\tonde{\sqrt{2}|\bar{\alpha}|\norm{\mb{v}}^2/\sqrt{\mb{v}^H\bs{\Gamma}\mb{v}},\sqrt{\lambda}}.
\ee

\bibliographystyle{IEEEtran}
\bibliography{ref_MIMO}

\begin{thebibliography}{10}
\providecommand{\url}[1]{#1}
\csname url@samestyle\endcsname
\providecommand{\newblock}{\relax}
\providecommand{\bibinfo}[2]{#2}
\providecommand{\BIBentrySTDinterwordspacing}{\spaceskip=0pt\relax}
\providecommand{\BIBentryALTinterwordstretchfactor}{4}
\providecommand{\BIBentryALTinterwordspacing}{\spaceskip=\fontdimen2\font plus
\BIBentryALTinterwordstretchfactor\fontdimen3\font minus
  \fontdimen4\font\relax}
\providecommand{\BIBforeignlanguage}[2]{{%
\expandafter\ifx\csname l@#1\endcsname\relax
\typeout{** WARNING: IEEEtran.bst: No hyphenation pattern has been}%
\typeout{** loaded for the language `#1'. Using the pattern for}%
\typeout{** the default language instead.}%
\else
\language=\csname l@#1\endcsname
\fi
#2}}
\providecommand{\BIBdecl}{\relax}
\BIBdecl

\bibitem{STAP}
W.~L. {Melvin}, ``A {STAP} overview,'' \emph{IEEE Aerospace and Electronic
  Systems Magazine}, vol.~19, no.~1, pp. 19--35, Jan 2004.

\bibitem{STAP_IID_ass_1}
E.~{Aboutanios} and B.~{Mulgrew}, ``Hybrid detection approach for {STAP} in
  heterogeneous clutter,'' \emph{IEEE Transactions on Aerospace and Electronic
  Systems}, vol.~46, no.~3, pp. 1021--1033, July 2010.

\bibitem{STAP_IID_ass_2}
A.~L. {Swindlehurst} and P.~{Stoica}, ``Maximum likelihood methods in radar
  array signal processing,'' \emph{Proceedings of the IEEE}, vol.~86, no.~2,
  pp. 421--441, Feb 1998.

\bibitem{Himed_GLRT}
K.~J. {Sohn}, H.~{Li}, and B.~{Himed}, ``Parametric {GLRT} for multichannel
  adaptive signal detection,'' \emph{IEEE Transactions on Signal Processing},
  vol.~55, no.~11, pp. 5351--5360, Nov 2007.

\bibitem{Ward}
J.~Ward, ``Space-time adaptive processing for airborne radar,''
  \emph{Massachusetts Institute of Technology, Lincoln Labs, Cambridge, MA,
  Tech. Rep. TR-1015, Dec. 1994}.

\bibitem{kay1993fundamentalsII}
S.~M. Kay, \emph{Fundamentals of statistical signal processing, volume II:
  detection theory}.\hskip 1em plus 0.5em minus 0.4em\relax Prentice Hall,
  1993.

\bibitem{marzetta2010noncooperative}
T.~L. Marzetta, ``Noncooperative cellular wireless with unlimited numbers of
  base station antennas,'' \emph{IEEE Trans. Wireless Commun.}, vol.~9, no.~11,
  pp. 3590--3600, Nov. 2010.

\bibitem{BjornsonHS17}
E.~Bj{\"o}rnson, J.~Hoydis, and L.~Sanguinetti, ``Massive {MIMO} has unlimited
  capacity,'' \emph{IEEE Trans. Wireless Commun.}, vol.~17, no.~1, pp.
  574--590, Jan. 2018.

\bibitem{massivemimobook}
\BIBentryALTinterwordspacing
E.~Bj\"{o}rnson, J.~Hoydis, and L.~Sanguinetti, ``Massive {MIMO} networks:
  {Spectral}, energy, and hardware efficiency,'' \emph{Foundations and
  Trends{\textregistered} in Signal Processing}, vol.~11, no. 3-4, pp.
  154--655, 2017. [Online]. Available:
  \url{http://dx.doi.org/10.1561/2000000093}
\BIBentrySTDinterwordspacing

\bibitem{Bjornson2019a}
\BIBentryALTinterwordspacing
E.~Bj\"ornson, L.~Sanguinetti, H.~Wymeersch, J.~Hoydis, and T.~L. Marzetta,
  ``Massive {MIMO} is a reality---{What} is next? {Five} promising research
  directions for antenna arrays,'' \emph{CoRR}, vol. abs/1902.07678, 2019.
  [Online]. Available: \url{http://arxiv.org/abs/1902.07678}
\BIBentrySTDinterwordspacing

\bibitem{Huber}
P.~J. Huber, ``The behavior of maximum likelihood estimates under nonstandard
  conditions,'' in \emph{Proceedings of the Fifth Berkeley Symposium on
  Mathematical Statistics and Probability, Volume 1: Statistics}.\hskip 1em
  plus 0.5em minus 0.4em\relax Berkeley, Calif.: University of California
  Press, 1967, pp. 221--233.

\bibitem{white}
H.~White, ``Maximum likelihood estimation of misspecified models,''
  \emph{Econometrica: Journal of the Econometric Society}, pp. 1 -- 25, 1982.

\bibitem{SPM}
S.~Fortunati, F.~Gini, M.~S. Greco, and C.~D. Richmond, ``Performance bounds
  for parameter estimation under misspecified models: Fundamental findings and
  applications,'' \emph{IEEE Signal Processing Magazine}, vol.~34, no.~6, pp.
  142--157, Nov 2017.

\bibitem{TSP_MCRB}
S.~Fortunati, F.~Gini, and M.~Greco, ``The {M}isspecified {C}ram{\'e}r-{R}ao
  bound and its application to scatter matrix estimation in complex
  elliptically symmetric distributions,'' \emph{IEEE Transactions on Signal
  Processing}, vol.~64, no.~9, pp. 2387 -- 2399, 2016.

\bibitem{Con_MCRB}
S.~Fortunati, F.~Gini, and M.~S. Greco, ``The constrained {M}isspecified
  {C}ram\'er-{R}ao bound,'' \emph{IEEE Signal Processing Letters}, vol.~23,
  no.~5, pp. 718--721, 2016.

\bibitem{miss_sb}
A.~Mennad, S.~Fortunati, M.~N.~E. Korso, A.~Younsi, A.~M. Zoubir, and
  A.~Renaux, ``Slepian-bangs-type formulas and the related misspecified
  {C}ram\'{e}r-{R}ao bounds for complex elliptically symmetric distributions,''
  \emph{Signal Processing}, vol. 142, pp. 320 -- 329, 2018.

\bibitem{white_nl_reg_dep}
H.~White and I.~Domowitz, ``Nonlinear regression with dependent observations,''
  \emph{Econometrica}, vol.~52, no.~1, pp. 143--161, 1984.

\bibitem{MMLE_dep}
I.~Domowitz and H.~White, ``Misspecified models with dependent observations,''
  \emph{Journal of Econometrics}, vol.~20, no.~1, pp. 35 -- 58, 1982.

\bibitem{icassp}
S.~{Fortunati}, L.~{Sanguinetti}, M.~S. {Greco}, and F.~{Gini}, ``Scaling up
  {MIMO} radar for target detection,'' in \emph{ICASSP 2019 - 2019 IEEE
  International Conference on Acoustics, Speech and Signal Processing
  (ICASSP)}, May 2019, pp. 4165--4169.

\bibitem{wide_MIMO}
A.~M. {Haimovich}, R.~S. {Blum}, and L.~J. {Cimini}, ``{MIMO} radar with widely
  separated antennas,'' \emph{IEEE Signal Processing Magazine}, vol.~25, no.~1,
  pp. 116--129, 2008.

\bibitem{Stoica_col}
J.~Li and P.~Stoica, ``{MIMO} radar with colocated antennas,'' \emph{IEEE
  Signal Processing Magazine}, vol.~24, no.~5, pp. 106--114, Sept 2007.

\bibitem{Fried_7}
B.~{Friedlander}, ``On signal models for {MIMO} radar,'' \emph{IEEE
  Transactions on Aerospace and Electronic Systems}, vol.~48, no.~4, pp.
  3655--3660, October 2012.

\bibitem{Abla}
A.~Kammoun, R.~Couillet, F.~Pascal, and M.~Alouini, ``Optimal design of the
  adaptive normalized matched filter detector using regularized {T}yler
  estimators,'' \emph{IEEE Transactions on Aerospace and Electronic Systems},
  vol.~54, no.~2, pp. 755--769, April 2018.

\bibitem{Large_scale}
H.~Jiang, Y.~Lu, and S.~Yao, ``Random matrix based method for joint {D}{O}{D}
  and {D}{O}{A} estimation for large scale {MIMO} radar in non-gaussian
  noise,'' in \emph{2016 IEEE International Conference on Acoustics, Speech and
  Signal Processing (ICASSP)}, March 2016, pp. 3031--3035.

\bibitem{Debbah}
P.~{Bianchi}, M.~{Debbah}, M.~{Maida}, and J.~{Najim}, ``Performance of
  statistical tests for single-source detection using random matrix theory,''
  \emph{IEEE Transactions on Information Theory}, vol.~57, no.~4, pp.
  2400--2419, April 2011.

\bibitem{kob}
H.~{Kobeissi}, Y.~{Nasser}, O.~{Bazzi}, A.~{Nafkha}, and Y.~{Lou\^{e}t},
  ``Elastic- enabling massive-antenna for joint spectrum sensing and sharing:
  How many antennas do we need?'' \emph{IEEE Transactions on Cognitive
  Communications and Networking}, vol.~5, no.~2, pp. 267--280, June 2019.

\bibitem{Stoica_MIMO_1}
L.~{Xu}, J.~{Li}, and P.~{Stoica}, ``Target detection and parameter estimation
  for {MIMO} radar systems,'' \emph{IEEE Transactions on Aerospace and
  Electronic Systems}, vol.~44, no.~3, pp. 927--939, July 2008.

\bibitem{Fried}
B.~Friedlander, ``On transmit beamforming for {MIMO} radar,'' \emph{IEEE
  Transactions on Aerospace and Electronic Systems}, vol.~48, no.~4, pp.
  3376--3388, October 2012.

\bibitem{libro_W}
F.~Gini, A.~{De Maio}, and L.~Patton, Eds., \emph{Waveform Design and Diversity
  for Advanced Radar Systems}, ser. Radar, Sonar \& Navigation.\hskip 1em plus
  0.5em minus 0.4em\relax Institution of Engineering and Technology, 2012.

\bibitem{MIMO_TAB}
I.~{Bekkerman} and J.~{Tabrikian}, ``Target detection and localization using
  {M}{I}{M}{O} radars and sonars,'' \emph{IEEE Transactions on Signal
  Processing}, vol.~54, no.~10, pp. 3873--3883, Oct 2006.

\bibitem{Stoica_book_MIMO}
J.~{Li} and P.~{Stoica}, \emph{{MIMO} Radar Signal Processing}.\hskip 1em plus
  0.5em minus 0.4em\relax Hoboken, NJ: Wiley, 2009.

\bibitem{Stoica_Q_2}
J.~{Li}, L.~{Xu}, P.~{Stoica}, K.~W. {Forsythe}, and D.~W. {Bliss}, ``Range
  compression and waveform optimization for {MIMO} radar: A cram\'{e}-rao bound
  based study,'' \emph{IEEE Transactions on Signal Processing}, vol.~56, no.~1,
  pp. 218--232, Jan 2008.

\bibitem{Fried_non_ort}
B.~{Friedlander}, ``Effects of model mismatch in {MIMO} radar,'' \emph{IEEE
  Transactions on Signal Processing}, vol.~60, no.~4, pp. 2071--2076, April
  2012.

\bibitem{bradley2005}
R.~C. Bradley, ``Basic properties of strong mixing conditions. {A} survey and
  some open questions,'' \emph{Probab. Surveys}, vol.~2, pp. 107--144, 2005.

\bibitem{book_CLT_dep_2}
T.~W. Anderson, \emph{An Introduction to Multivariate Statistical
  Analysis}.\hskip 1em plus 0.5em minus 0.4em\relax New York: John Wiley and
  Sons, Inc.,, 1958.

\bibitem{book_CLT_dep_1}
I.~A. Ibragimov and Y.~V. Linnik, \emph{Independent and Stationary Sequences of
  Random Variables}.\hskip 1em plus 0.5em minus 0.4em\relax The Netherlands:
  Wolters-Noordhoff, 1971.

\bibitem{CLT_dep}
M.~Rosenblatt, ``A central limit theorem and a strong mixing condition,''
  \emph{Proceedings of the National Academy of Sciences of the United States of
  America}, vol.~42, no.~1, pp. 43--47, 1956.

\bibitem{Pici}
B.~Picinbono, ``On circularity,'' \emph{IEEE Transactions on Signal
  Processing}, vol.~42, no.~12, pp. 3473--3482, 1994.

\bibitem{Pici_AR}
B.~Picinbono and P.~Bondon, ``Second-order statistics of complex signals,''
  \emph{IEEE Transactions on Signal Processing}, vol.~45, no.~2, pp. 411--420,
  Feb 1997.

\bibitem{Stoica_book}
P.~Stoica and R.~L. Moses, \emph{Spectral Analysis of Signals}, 2nd~ed., 2005.

\bibitem{CG_ran}
M.~{Rangaswamy}, D.~D. {Weiner}, and A.~{Ozturk}, ``Non-gaussian random vector
  identification using spherically invariant random processes,'' \emph{IEEE
  Transactions on Aerospace and Electronic Systems}, vol.~29, no.~1, pp.
  111--124, Jan 1993.

\bibitem{Sang}
K.~J. Sangston, F.~Gini, and M.~S. Greco, ``Coherent radar target detection in
  heavy-tailed compound-gaussian clutter,'' \emph{IEEE Transactions on
  Aerospace and Electronic Systems}, vol.~48, no.~1, pp. 64--77, Jan 2012.

\bibitem{MMLE_time_series}
T.~Bollerslev and J.~M. Wooldridge, ``Quasi-maximum likelihood estimation and
  inference in dynamic models with time-varying covariances,''
  \emph{Econometric Reviews}, vol.~11, no.~2, pp. 143--172, 1992.

\bibitem{Wald_de_maio_2}
A.~{De Maio}, S.~M. {Kay}, and A.~{Farina}, ``On the invariance, coincidence,
  and statistical equivalence of the {GLRT}, {R}ao test, and {W}ald test,''
  \emph{IEEE Transactions on Signal Processing}, vol.~58, no.~4, pp.
  1967--1979, April 2010.

\bibitem{Multivar_CLT_1}
A.~Bulinski and N.~Kryzhanovskaya, ``Convergence rate in {CLT} for
  vector-valued random fields with self-normalization,'' \emph{Probab. Math.
  Statist.}, vol.~26, no.~2, pp. 261--281, 2006.

\bibitem{White_B}
H.~White, ``Chapter {V}{I} - estimating asymptotic covariance matrices,'' in
  \emph{Asymptotic Theory for Econometricians}, ser. Economic Theory,
  Econometrics, and Mathematical Economics, H.~White, Ed.\hskip 1em plus 0.5em
  minus 0.4em\relax San Diego: Academic Press, 1984, pp. 132 -- 161.

\bibitem{CFAR_AMF}
F.~C. {Robey}, D.~R. {Fuhrmann}, E.~J. {Kelly}, and R.~{Nitzberg}, ``A cfar
  adaptive matched filter detector,'' \emph{IEEE Transactions on Aerospace and
  Electronic Systems}, vol.~28, no.~1, pp. 208--216, Jan 1992.

\bibitem{Complex_Normal}
N.~R. Goodman, ``Statistical analysis based on a certain multivariate complex
  gaussian distribution (an introduction),'' \emph{Ann. Math. Statist.},
  vol.~34, no.~1, pp. 152--177, 03 1963.

\bibitem{non_cenrtal_CDF}
A.~Nuttall, ``Some integrals involving the ${Q}_m$ function (corresp.),''
  \emph{IEEE Transactions on Information Theory}, vol.~21, no.~1, pp. 95--96,
  January 1975.

\bibitem{Stoica_ident}
J.~{Li}, P.~{Stoica}, L.~{Xu}, and W.~{Roberts}, ``On parameter identifiability
  of mimo radar,'' \emph{IEEE Signal Processing Letters}, vol.~14, no.~12, pp.
  968--971, Dec 2007.

\bibitem{ULLN}
B.~M. P\"{o}tscher and I.~R. Prucha, ``A uniform law of large numbers for
  dependent and heterogeneous data processes,'' \emph{Econometrica}, vol.~57,
  no.~3, pp. 675--683, 1989.

\bibitem{white_theo}
H.~White, ``Nonlinear regression on cross-section data,'' \emph{Econometrica},
  vol.~48, no.~3, pp. 721--746, 1980.

\bibitem{Cox}
D.~R. Cox and D.~V. Hinkley, \emph{Theoretical Statistics}.\hskip 1em plus
  0.5em minus 0.4em\relax Chapman and Hall/CRC, 1979.

\bibitem{Complex_ollila}
J.~{Eriksson}, E.~{Ollila}, and V.~{Koivunen}, ``Essential statistics and tools
  for complex random variables,'' \emph{IEEE Transactions on Signal
  Processing}, vol.~58, no.~10, pp. 5400--5408, Oct 2010.

\bibitem{Billi}
P.~Billingsley, \emph{Convergence of Probability Measures}, 2nd~ed.\hskip 1em
  plus 0.5em minus 0.4em\relax John Wiley \& Sons, 1999.

\end{thebibliography}

\end{document}